# Understanding Magnesium Dissolution through Machine Learning Molecular Dynamics

Zhoulin Liu[a], Jianchun Sha[b], Guang-Ling Song[c*], Ziliang Wang [d*], Yinghe Zhang[a*]

[a] School of Science, Harbin Institute of Technology, Shenzhen 518055, Guangdong, PR China

[b] Key Lab of Electromagnetic Processing of Materials, Ministry of Education, Northeastern University, Shenyang 110819, PR China

[c] Department of Ocean Science and Engineering, Southern University of Science and Technology, 1088 Xueyuan Boulevard, Shenzhen 518055, China

[d] National Engineering Laboratory for Reducing Emissions from Coal Combustion, School of Nuclear Science, Energy and Power Engineering, Shandong University, Jingshi Road, No. 17923, Jinan 250061, Shandong, China

*Corresponding authors. E-mail addresses: songgl@sustech.edu.cn (Guang-Ling Song), zwang2022@sdu.edu.cn (Ziliang Wang), zhangyinghe@hit.edu.cn (Yinghe Zhang).

**ABSTRACT:** Magnesium alloys have become increasingly important for various potential industrial applications, especially in energy storage, due to their outstanding properties. However, a clear understanding of the dissolution mechanism of magnesium in the most common aqueous environments remains a critical challenge, hindering the broader application of magnesium alloys. To address pending key controversies in magnesium alloys research, the atomic-scale hydrogen evolution process and dissolution mechanism of magnesium were investigated by combining machine learning molecular dynamics with density functional theory. These controversies include the presence of magnesium reaction intermediates, the formation of uni-positive $Mg^+$, the specific reaction steps involved in hydrogen evolution and magnesium dissolution, and the generation and growth mechanisms of the surface films. The results indicate that the intermediate species in the magnesium dissolution process is solid-phase MgOH, which exhibits an MgO-like structure. The magnesium in MgOH is identified as the widely recognized uni-positive $Mg^+$. The intermediate film is formed, consisting primarily of the MgOH phase with a small amount of MgO. This film grows inward by extending into the magnesium substrate. Under sufficient water availability, the film undergoes further oxidation to form $Mg(OH)_2$. These findings highlight the critical role of the MgOH phase in the magnesium dissolution process, leading to the proposal of a dissolution model based on MgOH/MgO solid phases as intermediates. These insights deepen the understanding of magnesium dissolution, pave the way for the development of more effective anti-corrosion strategies for magnesium alloys, and may also advance the utilization of magnesium in energy storage applications.

**Introduction**

Magnesium and its alloys have long been prominent in engineering applications and scientific research.[1,2] As the lightest structural metal, magnesium alloys exhibit significant potential in the automotive and aerospace industries.[3] Magnesium's capacity for safe degradation within the human body, combined with its minimal environmental impact, positions it as an ideal biodegradable metal, thereby granting it a unique role in the biomedical field.[4–6] Moreover, its low standard electrode potential and ability to transfer two electrons in chemical reactions have made it a focus of research in energy storage technologies, particularly in Mg-air batteries[7–9] and hydrogen storage[10] systems. Recently, it was proposed that Mg alloys could act as an intelligent anode to smartly monitor and adequately protect reinforced concrete from the corrosion attack caused by chloride ingress and carbonation.[11–14]

Previous studies have demonstrated that magnesium alloys exhibit relatively high corrosion rates in aqueous solutions. For instance, the AZ31 alloy[15] corrodes at a rate of 9.93 mm·year$^{-1}$ in a 0.15 M NaCl solution, and some magnesium alloys have even been reported to corrode at rates as high as 200 mm·year$^{-1}$.[8,16,17] This is unacceptable for practical applications. Furthermore, the application of magnesium alloys in the biomedical field, particularly for orthopedic implants, requires a controlled degradation rate to align with the healing rate of bone tissue.[18] Therefore, in-depth studies of the electrochemical dissolution behavior of magnesium alloys are critical for optimizing their biocompatibility and degradation performance.

Simultaneously, the development of Mg-air batteries and hydrogen storage technologies depends heavily on the electrochemical activity of magnesium,[19,20] which is closely linked to its dissolution behavior. Similarly, the electrochemical behavior of Mg closely influences the intelligence of an anode in protecting a reinforced concrete structure.[14] Thus, a comprehensive understanding of the dissolution mechanisms of magnesium is essential for improving the performance of these energy storage systems. Research into the dissolution process of magnesium has played an indispensable role in the historical development of the magnesium industry.

Song et al.[21] noted that a consensus has been reached regarding the dissolution of magnesium: the corrosion reaction of magnesium in aqueous solutions involves the electrochemical decomposition of water, producing hydrogen gas and magnesium hydroxide,

$$Mg + 2H_2O \rightarrow Mg^{2+} + 2OH^- + H_2 \tag{1}$$

or

$$Mg + 2H_2O \rightarrow Mg(OH)_2 + H_2 \tag{2}$$

in which there is

$$Mg^{2+} + 2OH^- = Mg(OH)_2 \tag{3}$$

The overall anodic dissolution is always accompanied by an overall cathodic hydrogen evolution, which can be expressed as:[22,23]

$$Mg \rightarrow Mg^{2+} + 2e^- \tag{4}$$

$$2H_2O + 2e^- \rightarrow 2OH^- + H_2 \tag{5}$$

In other words, the characteristic dissolution of Mg can be regarded as a hydrogen evolution corrosion process. Such a corrosion can also occur on the anode in a Mg-air battery. Moreover, oxygen reduction reaction (ORR) may also be involved in the cathode process due to the presence of oxygen in electrolytes, particularly in a Mg-air battery. The ORR can be expressed as follows:

$$H_2O + O_2 + 4e^- = 4OH^- \tag{6}$$

Correspondingly, the overall reaction of Mg corroded by oxygen is:

$$Mg + O_2 + H_2O = Mg(OH)_2 \qquad (7)$$

In current magnesium dissolution research, two widely cited models are the incomplete film uni-positive $Mg^+$ (IFUM) mechanism and the enhanced catalytic activity (ECA) mechanism.[21,24] The key perspectives of these models are as follows: IFUM suggests that during magnesium dissolution, magnesium first loses one electron to form a uni-positive $Mg^+$ containing intermediate that may be in various forms. The ECA mechanism proposes that a catalyst is present, which becomes more active with increasing potential to enhance the rate of the hydrogen evolution reaction (HER).

However, these models fail to adequately address below-mentioned key issues.[21,24] First, while both the IFUM and ECA mechanisms assert that the electrochemical reaction product of magnesium is $Mg(OH)_2$, experimental observations reveal a dense intermediate product film on the magnesium surface. The prevailing viewpoint is that this intermediate film consists of MgO[25,26]. This contradicts many existing magnesium dissolution mechanisms. For instance, both the ECA and IFUM mechanisms suggest that $Mg^{2+}$ is directly released from the magnesium substrate, with $Mg(OH)_2$ ultimately forming in the solution via reaction (3).[21] Despite this, the observation of intermediate products challenges the current electrochemical dissolution models, which fail to explain their formation in detail.

Second, although the IFUM model is widely used to explain the electrochemical reactions of magnesium, there is still a lack of robust theoretical and experimental evidence to support the presence and involvement of the $Mg^+$ containing intermediate in the anodic dissolution of Mg. As a result, validating this model through direct experimental observations or theoretical calculations remains a significant challenge in the field of magnesium dissolution.

In this work, a comprehensive analysis of hydrogen evolution, dissolution, and film growth processes is conducted during magnesium dissolution at the open-circuit potential (OCP) to address the key contentious issues in the field of magnesium research. These issues include the formation of uni-positive $Mg^+$, the detailed reaction steps of hydrogen evolution and magnesium dissolution, and the mechanisms of corrosion product film formation and growth. Machine learning molecular dynamics (MLMD) combined with density functional theory (DFT) is proposed to investigate the dissolution mechanisms of magnesium. The MLMD approach is applied to calculate the interactions between the magnesium surface and the solution and directly analyze the formation and growth mechanisms of intermediate product films in an aqueous environment. Based on these, a dedicated model is developed for magnesium dissolution at the OCP. The research provides not only key scientific insights into the understanding of Mg corrosion but also significant theoretical and practical implications for advancing magnesium applications in energy storage systems.

**Methods**

**DFT Calculation Details.** The electronic structure calculations were performed using the Vienna Ab initio Simulation Package (VASP).[27] The Perdew-Burke-Ernzerhof (PBE) generalized gradient approximation (GGA) was employed to describe the exchange-correlation energy.[28] The projector-augmented wave (PAW) method[29] was used to handle core-valence electron interactions, where the valence electrons for Mg ($3s^2$), O ($2s^2 2p^4$), and H ($1s^1$) were expanded on a plane-wave basis, setting the cutoff energy to 520 eV. A Monkhorst-Pack k-point grid centered at the Gamma point was used for Brillouin zone sampling.[30] For periodic structures, the k-point sampling interval was set to 0.03 Å$^{-1}$. The electronic level occupation was treated with a Gaussian smearing of 0.1 eV. Entropy corrections due to thermal occupation were subtracted from total energy calculations. The Grimme D3 empirical dispersion scheme was employed to account for van der Waals interactions.[31] The convergence criterion for electronic self-consistent interactions was set to $1\times10^{-5}$ eV·unit$^{-1}$. During the structural optimization process, the maximum force convergence criterion was set to 0.02 eV·Å$^{-1}$. To investigate ion transport properties, ab initio molecular dynamics (AIMD) simulations were combined with the Climbing Image-Nudged Elastic Band (CI-NEB) method.[32] Constant-potential molecular dynamics (CPMD) simulations were performed using the doped Ne electrode method developed by Surendralal et al.[33,34] The target temperature of the control system was 1000 K, with an initial control voltage of –1.0 V vs. OCP during the first 5 ps, followed by an increase to –1.5 V vs. OCP in the subsequent 5 ps. Further details on the computational methodology can be found in the "Calculation Methods" section of the Supporting Information.

**Construction of Training and Test Sets, and MLMD Simulations.** All machine learning potential (MLP) training and MLMD simulations were conducted using the GPUMD 3.9.5 software package,[35] with version 4 used for the training of the neuroevolution potential (NEP) model.[36] The overall MLP training process was divided into initial MLP training and active learning iterations. During the initial potential training, the initial training set was constructed through perturbation, AIMD simulations, empirical potential calculations, extraction from other training sets,[37,38] and a small number of manually constructed structures. After obtaining the structures, energy, force, and virial information were calculated using the VASP software.[27] During the active learning process, once the initial potential construction was completed, structures were generated through molecular dynamics simulations and sampled to improve the training set. During this process, the farthest-point sampling method[39] was used to obtain structures. The energy, force, and virial information were calculated using DFT and added to the training set. For the final 10 ns of calculations, sampling checks were carried out to ensure that the energy and force errors were consistent with those of the training set. Given the specificity of magnesium dissolution, separate training sets were constructed for the solution, interface, and substrate. Then, the separate training sets were later merged to produce the final training set. The test set was obtained during the training set construction and the final active learning sampling. In total, the training set contains 2902 structures and 253,031 atoms, while the test set contains 1224 structures and 86,387 atoms. Electronic structure and energy calculations were automatically performed using the VaspTool software package.[40]

After completing the training of the MLP, the MLMD simulations were performed. Before the simulations, the temperature was set to 300 K and the pressure to 1 atmosphere. The system was then relaxed under an isothermal-isobaric ensemble (NPT) to reduce internal stresses. Subsequently, the system was heated to the desired temperature and molecular dynamics simulations were conducted under the Langevin method in a canonical (NVT) ensemble.[41] The time step was set to 0.5 fs. Since dispersion interactions were not considered during the potential training, DFT-D3 corrections were applied during the molecular dynamics simulations.[42] The total simulation time was 5 ns, which is typically 100 times longer than the duration achievable by AIMD calculations. Details on the input parameters for the NEP training, DFT calculations, and MLMD simulations can be found in the Supporting Information "Input Parameters" section. The reactive force field[43] was used for comparison with the NEP model, and the relevant reactive force field calculation parameters and results are available in the Supporting Information under reactive force field calculations. Detailed files, including the NEP training set, NEP training parameters, VASP calculation parameters, MLMD model and calculation parameters, can be accessed via: https://github.com/linger1234567/Mg_corr_mech.

**Results and Discussion**

**NEP-MLP Model Training and Validation.** Figure 1a shows the loss function plot, where the loss function converges after approximately 100,000 iterations. The parity plots for energy, force, and stress confirm the high accuracy of the NEP model (Figures 1b-d). Generally, the presence of multiple elements and solid-liquid phases often reduces the fitting accuracy of a system. Despite these factors, the NEP model still performed well. The system exhibits a small root mean square error (RMSE), with values for energy, force, and stress being 13.124 meV·atom$^{-1}$, 181.537 meV·Å$^{-1}$, and 55.187 meV·atom$^{-1}$, respectively, for the training set. For the test set, these values are 11.907 meV·atom$^{-1}$, 155.853 meV·Å$^{-1}$, and 35.982 meV·atom$^{-1}$. Additionally, the sampling check results for each separate training set used in the training are shown in Supporting Information, Figure S4.

In the NEP training, the descriptor vector consists of multiple radial and angular components, with the radial function defined as a linear combination of several basis functions.[37] The descriptor components are invariant to the arrangement of the same type of atoms, and these descriptor coefficients are treated as trainable parameters, which is crucial for effectively distinguishing different atom pairs that contribute to the descriptors. The training data generation method relies on the chemical transferability embedded in the radial function equation. Therefore, we can demonstrate this feature through principal component analysis of the descriptor space (Figure 1e). The test set is sufficiently surrounded by the training set. The sheet-like, rather than discrete island-like regions, indicate that the training set adequately covers the configurations of the three elements: Mg, O, and H.

To further validate the accuracy of our model, we calculated the forces of the training set using the reactive force field[43] and compared them with the forces from our trained NEP-MLP (Figure 1f). The RMSE value for the reactive force field is 929.782 meV/Å, which is much higher than the value for the NEP-MLP (181.537 meV/Å), demonstrating that the NEP-MLP is significantly more accurate than traditional reactive force fields. Additionally, the NEP-MLP also offers an extremely high molecular dynamics calculation speed. Using the same computational resources and achieving comparable accuracy, NEP-MLP is approximately one million times faster than AIMD (Figure 1g), forming the foundation of our new findings. Notably, the accuracy of NEP-MLP is significantly higher than that of the reactive force field[43], and its computational speed is about 10 times faster than the reactive force field. As shown in Figure 1h, we also calculated the radial distribution function (RDF) for O-O pairs in water solution, and the RDF almost perfectly coincides with that obtained from AIMD calculations. Furthermore, the energy and forces for the recently proposed MgOH structure are predicted.[44] Although such a crystal structure does not exist in the training set, NEP still accurately predicts the energy and forces (Figures 1i, j). These validations confirm that the model possesses high reliability.

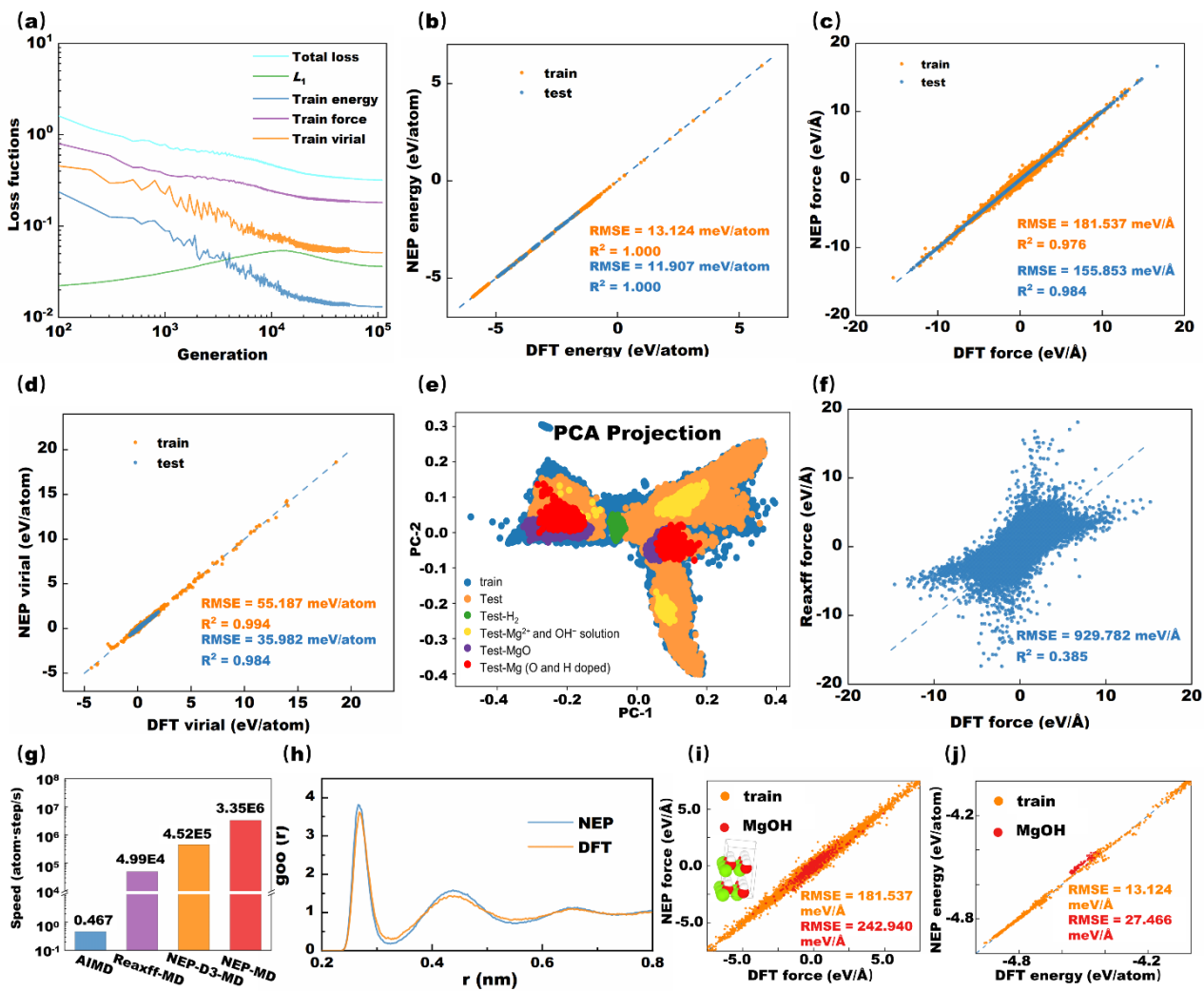

Figure 1. (a) Loss plot of NEP-MLP training; (b) Parity plot of energy for the NEP model; (c) Parity plot of force; (d) Parity plot of virial force; (e) PCA projection of the training set; (f) Parity plot of the reaction force field for NEP; (g) Comparison of computational speeds between AIMD, Reaxff-MD, and NEP-MD; (h) Comparison of radial distribution functions (RDFs) between NEP-MD and AIMD for aqueous solutions; (i) NEP prediction plot of forces for the MgOH structure, with the MgOH structure illustrated in the figure; (j) NEP prediction plot of energy for the MgOH structure.

**Multi-Stage Reaction Process on the Magnesium Surface.** The entire dissolution process of magnesium can be divided into several stages. In the early stage, the process begins with a smooth surface but ends with the surface being nearly entirely covered by hydroxide. In the second stage, the migration of OH∗ leads to the formation of the MgO/MgOH corrosion product film on the magnesium surface.

**Early Stage of Magnesium Dissolution.** The dissolution of magnesium on the surface involves a complex set of reactions, which are related to electron transfer. While the NEP-MLP model allows for long-duration and large-scale simulations, the current NEP-MD model cannot directly analyze charge transfer. However, the DFT calculations provide electronic structure information for the system. This enables us to analyze the reactions occurring on the magnesium surface in the early stage and better understand the charge transfer process during dissolution. Therefore, CPMD simulations of the electrode reactions were performed under the open-circuit condition (Figure 2a) or an applied potential (Figure 2b). The AIMD simulations in this paper are limited to very short times to reflect the potential reaction processes on the magnesium surface in the early stage. The electrode model, as shown in the ball-and-stick representation in Figures 2a and b, consists of magnesium atoms at the bottom representing the anode, Ne atoms at the top representing the cathode, and $H_2O$ molecules in the middle representing the electrolyte. The specific modeling and voltage simulation methods are described in the "Constant-potential molecular dynamics" section of the supporting information.

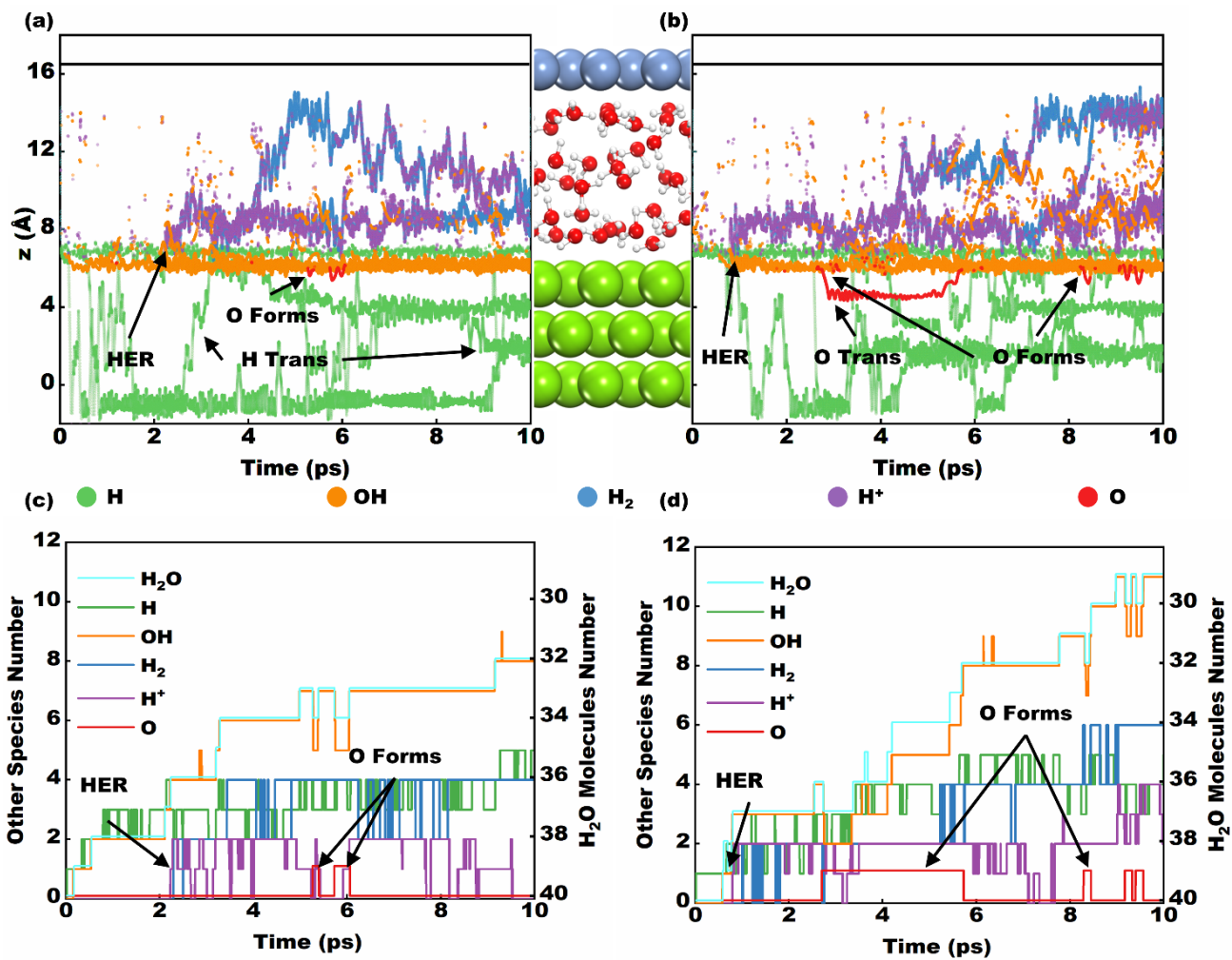

Figure 2. Temporal evolution of species formed under (a) the OCP and (b) the applied potential 1.0 V vs OCP during 0-5 ps and then 1.5 V vs OCP during 5-10 ps. Green, orange, blue, purple, and red dots represent adsorbed H* atoms, OH* (OH⁻ in solution), $H_2$, $H^+_{aq}$, and O*, respectively. In the electrolyte, atoms are shown as colored spheres: Mg (green), Ne (blue-purple), O (red), and H (white). The statistical number of species formed over time under (c) the OCP and (d) the applied potential 1.0 V vs OCP during 0-5 ps, and 1.5 V vs OCP during 5-10 ps, respectively. For clarity, the curves representing O* and remaining $H_2O$ molecules are offset upward by 0.1.

We first consider the case that Mg is at its OCP. It can be observed that the spontaneous dissociation of $H_2O$ molecules occurs rapidly (within 0.1 ps), and after the reaction, $H_{ads}$ migrates quickly between the Mg layers. This migration happens frequently. Nevertheless, the OH* tends to be immobilized on the surface upon formation. Around 2.1 ps, the HER occurs, and $H_2$ dissociates from the surface. The OH* count curve in Figure 2c shows that OH* forms rapidly in the first 5.0 ps, and the coverage number of OH∗ reaches 8 after another 5.0 ps (which corresponds to a coverage of approximately 2/3), eventually stabilizing.

To investigate the reactions under discharge conditions, a constant anodic potential is applied to the electrode surface (Figure 2b). The results show that the reaction rates of various processes are significantly accelerated during reaction (Figure 2c). The HER occurs as early as 0.85 ps. Moreover, MgO∗ formation is observed at 2.6 ps, followed by the inward migration of O∗ into the magnesium substrate.

However, unlike the open-circuit scenario, the coverage of OH∗ does not stabilize at 2/3 of a monolayer (Figure 2c). Instead, it continues to increase until nearly a complete monolayer is formed, consistent with structures observed in experimental studies.[45] Additionally, during the early stage of dissolution, the magnesium surface is entirely free of adsorbates, exhibiting high chemical reactivity. This suggests that some reactions occurring in this stage may not commonly appear in the later stage of magnesium dissolution.

To further clarify the redox reactions occurring in the early stage of magnesium dissolution, snapshots of the reactions are captured. Bader charge analyses are performed to study the associated charge transfer processes. The Bader charge values for the relevant atoms in each reaction are detailed in the Supporting Information (Tables S1-S5). The HERs observed here are consistent with those reported in the literature[46].

**Early stage of hydrogen evolution.** Magnesium dissolution is accompanied by hydrogen gas evolution. The hydrogen evolution process observed consists of two key steps: hydrogen storage on the magnesium surface, followed by hydrogen desorption from the surface. During hydrogen storage, $H_2O$ dissociates into OH∗ and H∗, as depicted in Figure 3a. This reaction can be expressed as:

$$H_2O \rightarrow OH* + H* \tag{8}$$

where ∗ denotes an adsorption site. This step appears to be a non-electrochemical reaction. However, Bader charge analysis (Supporting Information, Table S1) reveals that during this process, the magnesium substrate transfers one electron to H∗. Consequently, the reaction can also be represented as:

$$H_2O + e^- \rightarrow OH* + H^-* \tag{9}$$

The defining characteristic of this step is the acquisition of an electron and the production of OH*, marking it as an electrochemical step. This process also reserves OH* for subsequent reactions (as illustrated in Figure 3b). It is a Volmer-like hydrogen storage reaction, akin to reactions (8) and (9), where $H_2O$ gains an electron and dissociates into $OH^-$ and H*, as illustrated in Figure 3b,

$$H_2O + e^- \rightleftharpoons H* + OH^-_{aq} \quad (10)$$

where aq denotes the solvated species, with the resulting $OH^-_{aq}$ being dissolved in $H_2O$.

Once hydrogen storage progresses sufficiently, a second HER can occur. As shown in Figure 3c, this process involves the reaction between $H_2O$ and H*, leading to $H_2$ evolution.

This step is analogous to the Heyrovsky reaction[47] and can be expressed as:

$$H* + H_2O + e^- \rightleftharpoons OH^- + H_2 \quad (11)$$

Assuming the hydrogen evolution process follows reactions (8) or (10), followed by reaction (11), the calculated free energy diagram for this mechanism (see Supporting Information, Figure S10 and "Computational Hydrogen Electrode Method[48]" section) indicates that reaction (10) ($\Delta G = 0.88$ eV) is more difficult to occur compared to the other reactions. Therefore, the HER is divided into two parts, and the direct $H_2$ evolution reaction (11) does not occur independently. Instead, it requires the intermediate hydrogen storage on the surface, i.e., reaction (8) or (10), and the subsequent hydrogen evolution (11). Furthermore, the amount of hydrogen stored on the surface influences the probability of $H_2$ evolution. As a result, hydrogen evolution only occurs once the surface hydrogen storage process has progressed sufficiently. The HER in this context is consistent with the Volmer-Heyrovsky mechanism rather than the Tafel mechanism. This distinction arises from the formation of unique adsorption structures by anions on the magnesium surface, which facilitate the migration of adsorbed H* into the substrate, thereby hindering direct interaction between magnesium atoms and hydrogen.

It is worthwhile to stress that the cathodic hydrogen evolution takes place while the Mg surface is being corroded. Thus, the intermediates and final products from the cathodic reactions and anodic dissolution on the Mg surface may more or less interact with each other and influence the corresponding cathodic and anodic processes, and thus the overall corrosion damage of Mg.

**Early Stage of Magnesium Oxidation.** To investigate the oxidation of magnesium, we carefully examined the simulated trajectories related to magnesium dissolution reactions. When oxygen reduction dominates the cathodic process (with a high concentration of OH⁻ near the anode), the magnesium oxidation process can be divided into two steps. During the first step, OH* species gradually cover the Mg surface. As shown in Figure 3f, OH⁻ ions often aggregate within a 3-4 Å range from the surface, forming a solvation shell structure centered on OH⁻ ions. These OH⁻ ions may originate from the hydrogen storage reaction (Equation 10), HER (Equation 11), or ORR (Equation 6). This solvation shell structure promotes surface attack by facilitating electron transfer through the central OH⁻ ion or water molecules in the first solvation shell, as illustrated in Figure 3f. This electron transfer mechanism resembles the Grotthuss mechanism for proton transport. The electron-mediated transfer of OH⁻ significantly accelerates its diffusion compared to the diffusion of water molecules via Brownian motion.[49] Consequently, reactions involving OH⁻ on the magnesium surface occur much more rapidly. Therefore, suppressing this process is critical for mitigating excessive hydrogen evolution.[50–52] This process can be represented as:

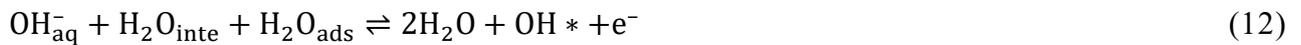

$$OH^-_{aq} + H_2O_{inte} + H_2O_{ads} \rightleftharpoons 2H_2O + OH* + e^- \qquad (12)$$

where $H_2O_{inte}$ denotes an intermediate water molecule involved in electron transfer. It describes the process in which a solvation shell structure with central OH⁻ gives out an electron to the substrate Mg and combines with an adsorbed water molecule to form 2 free water molecules away from the surface, leaving an OH* adsorbed on the substrate. It can be rewritten as a simplified anodic electrochemical adsorption of OH⁻ onto the magnesium substrate:

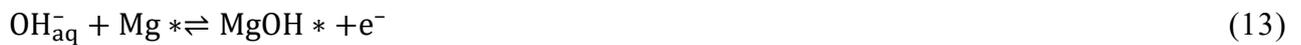

$$OH^-_{aq} + Mg* \rightleftharpoons MgOH* + e^- \qquad (13)$$

In this equation, Mg* represents a surface site, and the process is electrochemical, with the negative charge of OH⁻ transferring to the magnesium substrate.

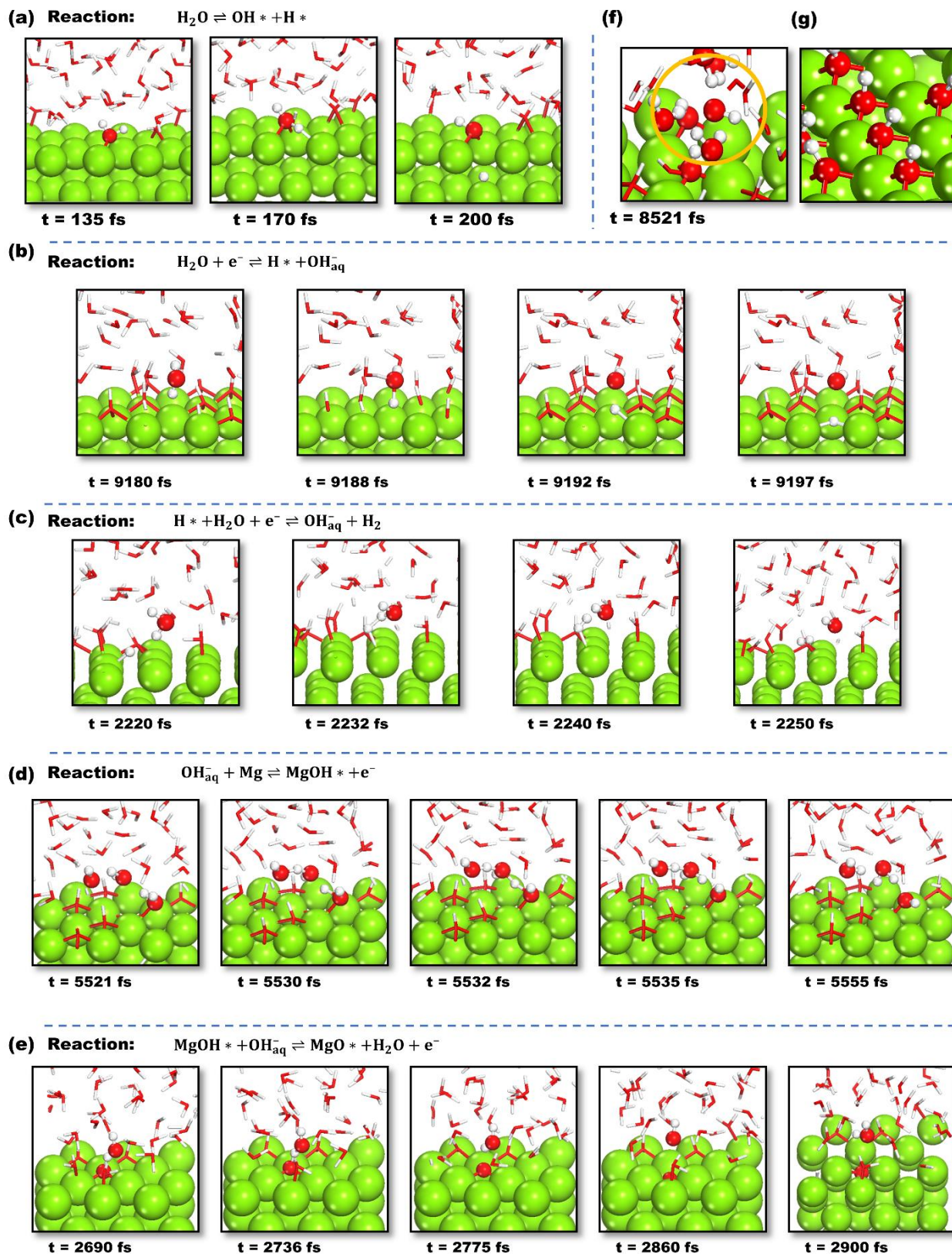

Figure 3. Reaction snapshots: (a) at the OCP during 135-200 fs, $H_2O$ dissociates into $H_*$ and $OH_*$, (b) at the OCP during 9180-9197 fs, $H_2O$ dissociates into $H_{ads}$ and $OH^-_{ads}$, (c) at the OCP during 2220-2250 fs, $H_2O$ reacts with $H_{ads}$ to produce hydrogen gas and $OH^-_{ads}$, (d) at the applied potential 1.5V vs. OCP during 5521-5555 fs, $OH^-$ reacts with magnesium to form $MgOH_*$, (e) at the applied potential

1.0V vs. OCP during 2690-2900 fs, OH⁻_aq reacts with MgOH* to form MgO*, and (f) a solvation shell composed of OH⁻ and H₂O.(g) the adsorption structure of hydroxide at the hollow site in three neighboring Mg atoms. Atoms are shown as spheres (Mg: green, O: red, H: white); less important atoms are depicted as stick models of the same color.

The OH⁻_aq ions in this reaction are derived from either water dissociation or, under practical conditions, the ORR described by Reaction (6). Assuming the OH⁻ ions are supplied predominantly by ORR (as occurs in Mg-air battery dissolution), the electrochemical reaction rate is faster than the water dissociation rate. Consequently, the majority of surface-adsorbed OH⁻ ions are provided by ORR (6), as shown in Figure 3d. This process can be expressed as:

$$Mg_n + OH^-_{aq} \rightleftharpoons Mg_nOH* + e^- \tag{14}$$

$$Mg_nOH* + OH^-_{aq} \rightleftharpoons Mg_n(OH)_2* + e^- \tag{15}$$

where n represents the number of surface magnesium atoms. As the reaction continues, the surface becomes fully covered:

$$Mg_n(OH)_{n-1}* + OH^-_{aq} \rightleftharpoons Mg_n(OH)_n* + e^- \tag{16}$$

This stepwise OH⁻ adsorption and electron transfer process resembles the hydroxide-assisted metal dissolution mechanism reported in the literature[53,54] except for some critical differences in the subsequent stages.

In the hydroxide-assisted dissolution mechanism, further OH⁻ adsorption is necessary:

$$Mg_n(OH)_n* + OH^-_{aq} \rightleftharpoons Mg_n(OH)_{n+1}* + e^- \tag{17}$$

And the Magnesium hydroxide may also be dissolved into the solution:

$$Mg_n(OH)_{n+1}* \rightleftharpoons Mg_{n-1}(OH)_{n-1}* + Mg^{2+}_{aq} + 2OH^-_{aq} \tag{18}$$

Then, the dissolved magnesium ions combine with OH⁻ ions to form magnesium hydroxide, as illustrated in Reaction (3).

However, the dissolution of the bulk phase requires overcoming a substantial energy barrier. In the present mechanism, once the surface is fully covered with OH*, the valence of surface magnesium atoms increases to +1. Subsequently, surface-adsorbed OH* reacts further with $OH^-_{aq}$, oxidizing the surface magnesium atoms to the higher valence +2, as illustrated in Figure 3e:

$$MgOH* + OH^-_{aq} \rightleftharpoons MgO* + H_2O + e^- \tag{19}$$

It should be noted that the MgO and MgOH are regarded as intermediate products. Due to the limitations of the AIMD scale in the early stage of Mg dissolution, the MgOH phase was not observed. However, it is a predominant intermediate during the second stage, which will be discussed in detail in the subsequent section.

It is interesting according to the computation that the surface oxidation state of magnesium can be a non-integer. Figure 4 illustrates the variation in the surface Mg oxidation state over time. The oxidation state rapidly increases in the period from 0 to 5 ps and then stabilizes, reaching an approximate +1 valence at around 10 ps (Figure 4a). The numerical difference in the surface magnesium atomic oxidation state is minimal between the open-circuit and applied potential scenarios, consistent with other studies on the effects of electric fields on oxidation states.[55] Additionally, Figure 4c shows that at the onset of the reaction, the labeled atoms have an oxidation state of approximately 0. As the reaction progresses, the oxidation state of surface atoms gradually shifts towards +1. Due to the unique adsorption structure of the $OH^-$ at the hollow site in three neighbor Mg atoms on the magnesium surface (Figure 3g), as well as the high conductivity within the metallic substrate, the charge is dispersed throughout the Mg substrate surface rather than being localized on a single atom. Consequently, when an $OH^-$ species is adsorbed onto the surface through an electrochemical reaction, it is not a single Mg atom that loses an electron, but rather three Mg atoms in direct contact that collectively lose one electron. The Figure also shows that the change in the surface Mg oxidation state from 0 to +1 is also a gradual process (Figure 4c-m). Since adsorption progresses, the oxidation of the surface Mg atoms is gradually shared among the neighboring Mg atoms until a monolayer is formed, with Mg atoms reaching the +1 oxidation state.

Similarly, for the second electron transfer reaction, although two electrons are transferred locally, a single magnesium atom still does not reach a +2 valence. This result is consistent with what is reported in some literature that the formation of a +2 valence is much more difficult than the +1 valence process.[56] Therefore, we propose that there are intermediate states between +1 and +2 valences, which significantly lower the energy barrier for the reaction from +1 to +2 valences. This can be evidenced by the structure of O/OH groups entering the Mg surface and the MgOH structure mentioned later, where O or OH groups always interact with multiple Mg atoms. In this case, the +1 valence can be considered as an intermediate process in the gradual change from 0 to +2 valence. The varying valence of Mg atoms on the surface embodies the $Mg^+$ containing species involved in the anodic dissolution of Mg in the IFUM model.[21]

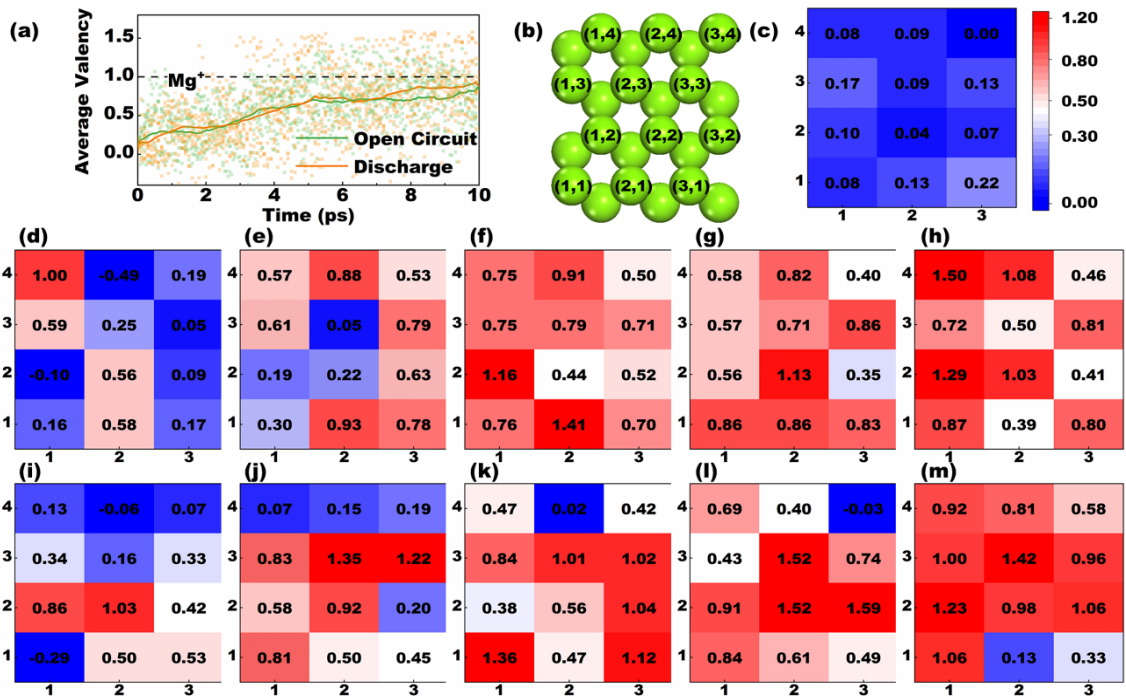

Figure 4. The valence state of surface Mg under open-circuit and applied potential conditions: (a) average valence state, with points in the background representing the valence state of each atom at different time steps, (b) surface Mg atom numbering, (c) initial state of surface Mg valence, (d)-(h) surface valence states in the period of 2-10 ps at the OCP, and (i)-(m) surface valence states at 2, 4, 6, 8, and 10 ps, respectively under applied potential conditions.

Figure 3e reveals the O atom spontaneously diffuses from the surface to the interior after its generation. To further deepen the understanding of the surface MgO formation mechanism, we used the CI-NEB method to calculate the energy barrier during the diffusion process from the Mg surface to analyze the role of this diffusion in the formation of MgO. Figure 5a illustrates the low-energy configurations that may form after oxygen diffusion into the surface, where the O1 structure corresponds to the situation in Figure 3e where O diffuses into the magnesium layer. There are six low-energy structures on the magnesium surface. To investigate the cause of the O atom's diffusion towards the interior, we calculated the energy change in this process (Figure 5b). The results indicate that the diffusion of the O atoms to the interior is a spontaneous process energetically, with no significant energy barrier, and as long as an $O_*$ is produced on the surface, it will tend to diffuse to the O1 site.

To explore whether the O atom at O1 tends to further move into the interior magnesium atomic layer, we calculated the energy barrier for its diffusion from O1 to the second and third interlayer gap sites. As shown in Figure 5c, among the three possible diffusion paths considered, the path from O1 through O3 to O6 in the second layer has the lowest activation energy, only 0.61 eV (Figure 5c). This lower barrier indicates that the diffusion of oxygen atoms within the magnesium atomic layer is a relatively easy process. For reference, the energy barriers of 0.1-0.5 eV are usually required for a common process in lithium-ion battery research, and thus a barrier of 0.82 eV can be considered to be relatively low (Supporting Information Table S6).[57] When different electric field strengths are applied to the Mg, the strength does not affect the diffusion of O in the Mg (Figure 5d). Due to the small model size, a larger energy barrier inside Mg is still exhibited.[58] Since Mg is a good electric conductor, the effect of the electric field is negligible in Mg. Therefore, the electric field is mainly distributed across the interface between Mg and the solution, where it regulates the electrochemical reaction by affecting the structure of the electric double layer in the electrolyte[59] or the reaction energy barrier at the interface[60]. However, in experiment, it does have been observed that the surface hydroxide/oxide layer increases evidently with increasing potential.[61] Hence, this may be attributed to the higher concentration of $O_*$ stored on the Mg surface resulting from the reactions (12) through (19) accelerated at a higher potential. It is the higher surface concentration, rather than the lower barrier in the Mg lattice that enhances the diffusion of O inward the Mg. Additionally, Figure 5e shows that when the concentration of oxygen is high, e.g., in case the MgO is formed, the diffusion energy barrier for the oxygen atom will be as high as 4.03 eV, indicating that under such a condition, the diffusion of oxygen atoms bonded in the Mg lattice becomes very difficult, implying that the MgO is a good barrier in the corrosion of Mg.

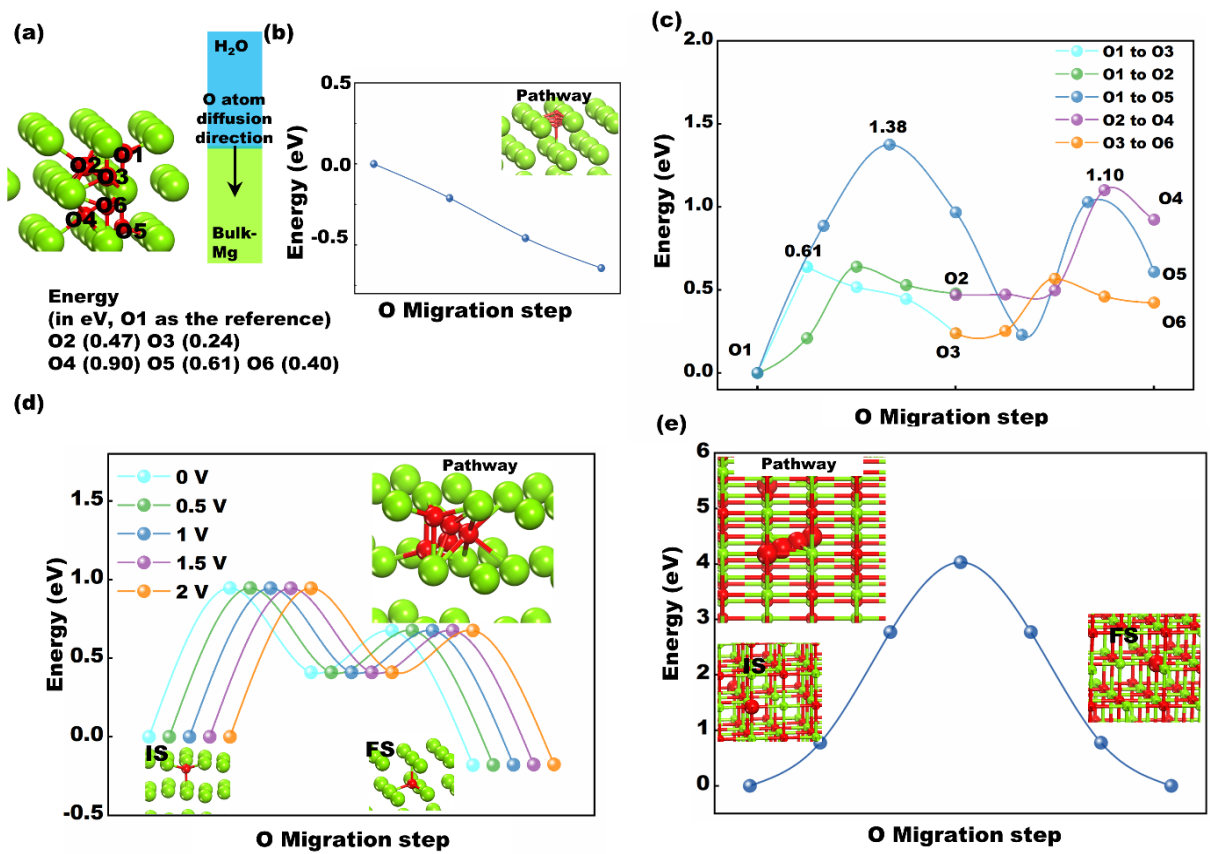

Figure 5. Stable structures and diffusion energy barriers of O on the Mg surface: (a)-(e); (a) stable structure of O atoms on the surface of Mg (b) diffusion steps and corresponding required energies of an O atom from the surface to the internal O1 site, (c) diffusion energy barriers of an O atom from the site to site: Path 1: from O1 to O5; Path 2: from O1 to O2, then to O4; Path 3: from O1 to O3, then to O6; diffusion energy barrier values are indicated in the figure, (d) diffusion energy barriers for O from O1 to O3 under different voltages, considering rate-determining step and curve shifting, and (e) diffusion energy barriers of O in Mg with high O concentration (represented as MgO); Atoms are shown as spheres (Mg: green, O: red, H: white); less important atoms are depicted as stick models of the same color.

**The Second Stage of Magnesium Dissolution.** Limited by the timescale achievable by AIMD simulations, the MLMD is employed to simulate the second stage of dissolution. First, we use small-scale (approximately 300 atoms) models to study the chemical reactions and intermediates during the magnesium dissolution process. Second, a larger-scale model (approximately 3400 atoms) is used to investigate surface cracking and the transformation of the intermediate MgOH phase to $Mg(OH)_2$/MgO.

**Chemical Reactions and Intermediates in the Second Stage of Magnesium Dissolution.** Figures 6a and b show the small-scale structures used, with Figure 6a depicting the close-packed (001) surface of magnesium, which is generally considered to have a lower dissolution rate.[62] A stepped surface consists of atomic terraces separated by steps, where Mg atoms at the steps have a lower coordination number, indicating high reactivity. We used the stepped surface in Figure 6b to describe the dissolution rate of high-index surfaces.

The oxidation of magnesium occurs directly on the magnesium substrate, progressing gradually from the outer layer to the inner layer (Figure 6e-f). In addition, there is a distinct difference in the dissolution behavior of magnesium on the close-packed and stepped surfaces. In the initial 200 ps (Figure 6c), $H_2O$ rapidly decomposes, forming adsorbed $OH_*$ and $H_*$. The $OH_*$ gradually covers the Mg surface, while adsorbed $H_*$ quickly diffuses inward the magnesium. After 200 ps, the surface reaches a steady state where $H_2O$ consumption becomes negligible, and $OH_*$ temporarily passivates the Mg surface and prevents further corrosion. Nevertheless, this passivation lasts only about 1200 ps, after which $H_2O$ consumption begins to gradually increase again. In contrast, on the stepped surface, after the rapid reaction in the initial 200 ps, there is only a steady state period of about 200 ps before $H_2O$ consumption rises quickly again (Figure 6d).

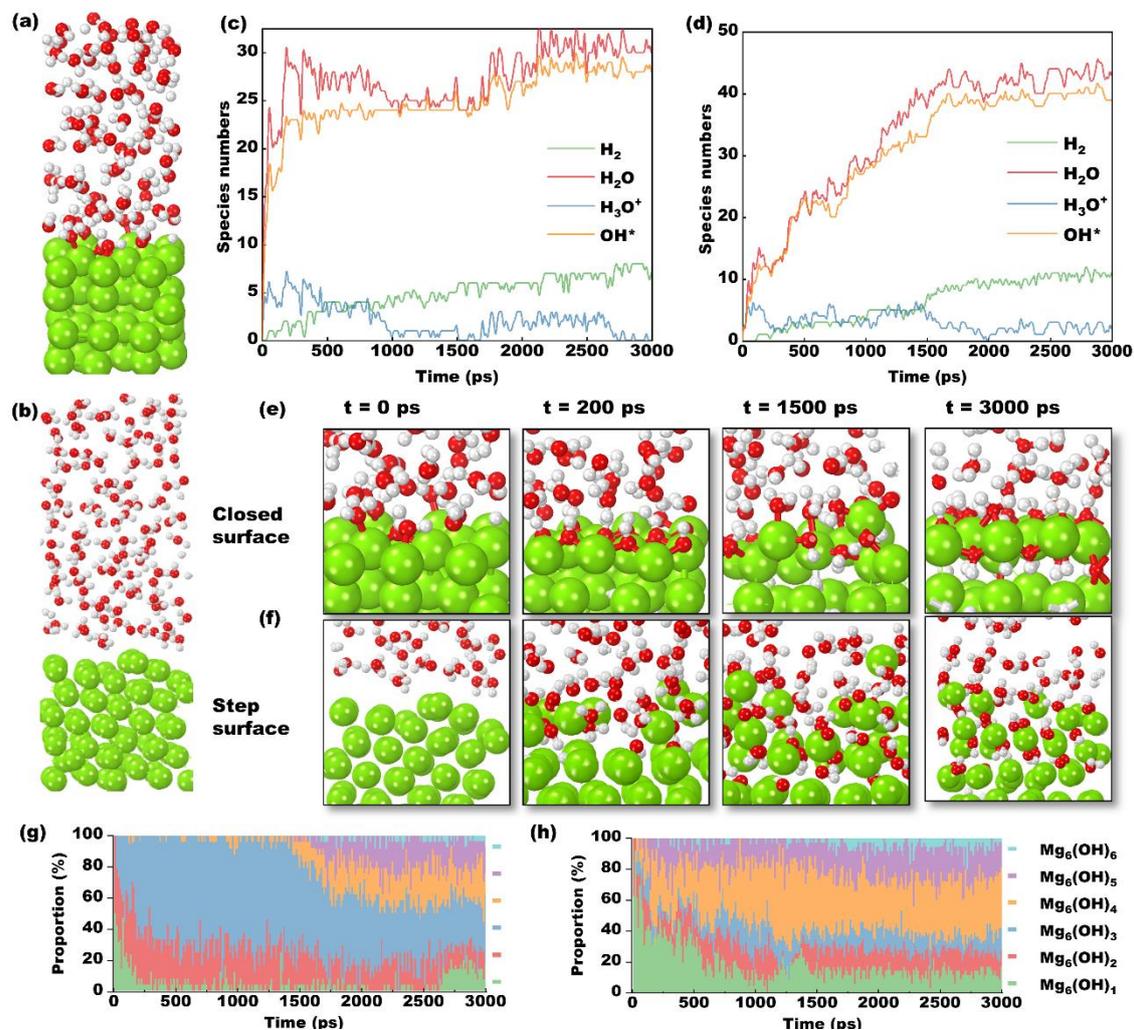

Figure 6 Results of the small model MLMD: (a) initial close-packed surface (0001) structure of Mg, (b) initial stepped surface structure of Mg, (c) species count versus time plot for reaction on the close-packed surface of Mg at 700K, (d) species count versus time plot for reaction on the stepped surface of Mg at 700K, (e) reaction snapshots on the close-packed surface over 0-3000ps, (f) reaction snapshots on the stepped surface over 0-3000ps, (g) the proportion of the magnesium oxide species versus time on the close-packed surface over 0-3000ps, and (h) the proportion of magnesium oxide species versus time on the stepped surface over 0-3000ps. Atoms are shown as spheres (Mg: green, O: red, H: white).

Figures 6e-f provide reaction snapshots at corresponding times. On the stepped surface, OH∗ diffuses inward the inner layer faster than on the close-packed surface. In the final snapshot (t = 3000 ps), a layered MgOH structure can be observed. The layered MgOH structure is supported by AIMD results, see Figure S12 of Supporting Information. Figures 6g-h show that on the close-packed surface, the Mg oxide species (MgOH$_n$) has a low OH$_*$ coordination number (1-3) initially, which constitutes a large proportion of the Mg oxide species on the surface. From 200 to 1500 ps, the Mg oxide species with 3-coordination OH$_*$ becomes predominant. After that the number of coordination OH$_*$ increases gradually (4-6). Contrarily, on the stepped surface, the Mg oxide species with 3-OH$_*$ coordination never predominate. Since the Mg oxide species with a small number of OH$_*$ quickly transforms into multi-OH$_*$ coordinated magnesium.

To investigate the transformation of MgOH$_*$ and the mechanism of hydrogen evolution, snapshots of the corresponding reaction processes are identified in the trajectory. Figure 7 reveals the Mg surface fully covered by OH$_*$, on which H$_2$O molecules can still be adsorbed onto magnesium atoms. As the magnesium adsorbed by OH$_*$ gradually moves inward from the surface under the influence of H$_2$O, the adsorbed OH$_*$ migrates from the surface inward along the gaps among the Mg atoms. After this migration step, a vacancy on the magnesium surface is formed, allowing for the reaction of water from the electrolyte or the adsorption of OH$^-$ ions. Repeatedly, a MgOH layer is gradually formed on the surface.

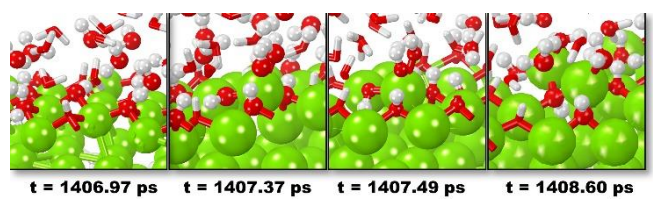

Figure 7. Mechanistic illustration of the migration of OH$_*$ from the surface inward to the inner layers to form MgOH phase. Atoms are shown as spheres (Mg: green, O: red, H: white); less important atoms are depicted as stick models of the same color.

Since the OH$_*$ can occupy the magnesium vacancy on the surface, resulting in surface passivation and reducing the magnesium surface reactivity, it will be difficult for H$_*$ on the magnesium surface to migrate and combine with the adsorbed H$_2$O on the surface to form H$_2$. As the Gibbs free energy of H$_2$O adsorption on the surface is negative, water can be spontaneously adsorbed and dissociated on the Mg. On the Mg surface nearly saturated with OH$_*$ adsorption, an anodic H$_2$O adsorption and cathodic hydrogen evolution are observed, as shown in Figure 8. A water molecule, assisted by another water molecule, undergoes dissociative adsorption onto the surface. However, due to the isolation by surface OH*, H is difficult to be adsorbed onto the magnesium substrate. Thus, it is combined with another water molecule to form a hydronium ion ($H_3O^+$). Subsequently, the hydronium ion undergoes dissociation to form $H_2O$ and solvated $H_{aq}$. Bader charge calculations indicate that the $H_{aq}$ is of 0 valence.

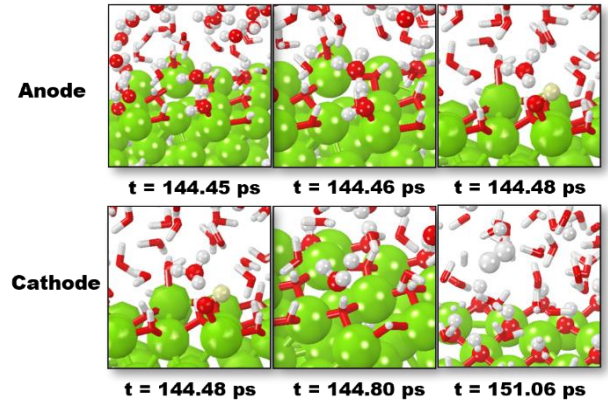

Figure 8. Schematic illustration of the anodic generation of $H_3O^+$ and the cathodic hydrogen evolution. Atoms are shown as spheres (Mg: green, O: red, H: white); less important atoms are depicted as stick models of the same color.

To understand the anodic and cathodic processes illustrated in Figure 8, two possible theoretical interpretations are proposed:

The first one is based on the following anodic process of

$$Mg + 2H_2O \rightarrow MgOH* + H_3O^+ + e^- \tag{20a}$$

where the MgOH* is actually the uni-positive $Mg^+$ containing species that has been widely proposed,[21] which may be further decomposed to $Mg^+$:

$$MgOH* \rightleftharpoons Mg^+ + OH^- \tag{20b}$$

Hence, reaction (20a) can be represented as:

$$Mg + 2H_2O \rightarrow [Mg^+ + OH^-] + H_3O^+ + e^- \tag{21}$$

Meanwhile, there is a cathodic reaction on the Mg surface to convert the generated hydronium ion from reaction (20a) to solvated $H_{aq}$:

$$H_3O^+ + e^- \rightarrow H_2O + H_{aq} \tag{22}$$

Subsequently, two $H_{aq}$ atoms combine together to form $H_2$:

$$H_{aq} + H_{aq} \to H_2 \tag{23}$$

In this explanation, the first reaction (20a) or (21) is an anodic reaction, whereas the second one (22) is a cathodic process. As the applied potential increases, the energy required for the first anodic reaction decreases (see Supporting Information Figure S11), and its rate increases, thus increasing the concentration of hydronium in the solution. As hydronium can diffuse away very fast, only a very limited proportion of the generated hydronium ions in the solution participated in a cathodic reaction (22) on the electrode surface. Meanwhile, due to the retroactive effect of the potential, reaction (22) cannot be accelerated significantly, and thus hydrogen evolution (23) cannot always increase with increasing potential. One may argue that the real potential actually does not increase (the real potential is fixed) due to the increasing IR drop when the applied potential positively shifts, and hence the cathodic hydrogen evolution rate can still keep increasing due to the increasingly roughened Mg surface under anodic polarization. If this is the case, then the anodic current density must increase equally with the hydrogen evolution rate. Unfortunately, in experiment, the anodic current density always increases much faster than the hydrogen evolution as the applied potential increases.[23] Moreover, according to reactions (21) and (22), the concentration of hydronium will increase (i.e., the solution is being acidified) with increasing potential and time. This also contradicts the solution alkalization results in experiment.[61,63]

In theory, reactions (20a) or (21) are more likely to take place on metallic Mg that is not covered by OH* or O (i.e., the Mg has not been oxidized to MgOH* or MgO). Since most of the Mg surface has been covered by OH and O in the early stage, the reaction (20a) or (21) in practice is obviously a low-possibility event. Also, reaction (22) is in nature a Volmer process (10), more likely occurring on metallic Mg, rather than on the surface covered by OH and O. Furthermore, since it has been computerized that the Volmer reaction (22) has a higher energy barrier than the Heyrovsky reaction (11) (see Figure S10), the latter is not the dominating cathodic process in practice where sufficient H* atoms are already present on the Mg surface.

The second possible interpretation for Figure 8 in theory is based on the surface that has experienced the early stage of oxidation. That is, Mg has been covered by the MgOH* formed via reaction (13) in the early stage of Mg oxidation. The MgOH∗ as an intermediate on the surface can be further chemically oxidized to Mg(OH)$_2$:

$$MgOH* + H_2O \rightleftharpoons Mg(OH)_2 + H_{ad} \tag{24a}$$

The formed Mg(OH)$_2$ may be further decomposed into Mg$^{2+}$:

$$Mg(OH)_2 \rightleftharpoons Mg^{2+} + 2OH^- \tag{24b}$$

Hence, reaction (24a) can be rewritten into

$$MgOH* + 2H_2O \rightarrow [Mg^{2+} + 2OH^-] + H_{ad} \tag{25}$$

It can also be regarded as a combination of an anodic process and a cathodic process:

$$MgOH* \rightleftharpoons MgOH*^+ + e^- \tag{25a}$$

$$MgOH*^+ + H_2O + e^- \rightleftharpoons [Mg^{2+} + 2OH^-] + H_{ad} \tag{25b}$$

They occur almost the same time. The generation of OH$^-$ in reaction (25) leads to the well-known Mg surface alkalization effect.[62] The neighbor adsorbed H$_{ad}$ atoms on the surface combines together to form hydrogen gas H$_2$:

$$H_{ad} + H_{ad} \rightarrow H_2 \tag{26a}$$

They may also be desorbed into the solution, solvated as H$_{aq}$, then combined with the adsorbed H$_{ad}$ atoms to form H$_2$, and finally released from the surface:

$$H_{aq} + H_{ad} \rightarrow H_2 \tag{26b}$$

As the concentration of the intermediate MgOH∗ resulting from reaction (13) increases with increasing potential, the anodic hydrogen evolution process (26 a and b) will be accelerated by anodic polarization. Since on the electrode surface, there is still a limited metallic Mg area not covered by OH∗ and O, cathodic hydrogen evolution (10) and (11) may take place:

$$H_2O + e^- \rightleftharpoons H* + OH^-_{aq} \tag{27}$$

$$H* + H_2O + e^- \rightleftharpoons OH^- + H_2 \tag{28}$$

However, with increasing potential, the cathodic hydrogen evolution will slow down.

Compared with the first interpretation, the dissolution of Mg (25) and the hydrogen evolution (26 a and b) in the second interpretation mainly take place in the area covered by intermediate OH$_*$. As the OH$_*$ covered area is reasonably larger than the metallic Mg area, the reactions in the second interpretation should be more likely to occur than those in the first interpretation. Moreover, both the anodic hydrogen resulting from the combination of H$_{ad}$ and/or H$_{aq}$ atoms (26 and b) and the cathodic hydrogen (27) and (28) in the second interpretation are released from the surface, while the cathodic hydrogen combined of H$_{aq}$ in the first interpretation is released from the solution rather than the electrode surface. The former is a common experimental phenomenon, while the latter has not been reported. In summary, the first interpretation may describe the anodic dissolution and cathodic hydrogen evolution processes in the tiny area where metallic Mg is directly exposed to the solution, while the second interpretation presents the anodic processes in the area where the Mg has been oxidized to an intermediate.

**Surface Cracking and the Formation and Transformation of MgOH Intermediates.** The larger-scale and longer-duration MLMD simulations are conducted to eliminate size effects and better investigate surface cracking, the structure of the intermediate MgOH phase, and its transformation to Mg(OH)$_2$. Figures 9d and l show the close-packed surface and stepped surface models respectively used in this paper. Regarding the OH$_*$ coverage on the close-packed surface, previous thermodynamic studies have shown that the adsorption of OH$_*$ on the magnesium surface is always spontaneous until full coverage is reached.[53,54,64] However, in our previous AIMD calculations of the early magnesium dissolution, as well as in some studies,[33] the surface OH$_*$ coverage never reached the full coverage which is expected by thermodynamic calculations. Even when the temperature was increased up to 450K, the coverage only balanced at 1/4. Therefore, some researchers have considered various coverages,[54] or directly assumed coverage of 1 for calculations.[46] Here, thanks to the speed advantage of the MLMD, we are able to directly confirm that under conditions of 350K, the coverage reached nearly 1 (see Figures 9a and e) (the scale achievable by AIMD is indicated).

To directly obtain the corroded surface of magnesium, we appropriately increase the temperature of the system to accelerate the reaction process. From Figure 9b, it can be observed that the reaction rate for the large system is significantly faster than that for the small system. This can be explained by the size effect in small systems, where the distance between periodic images is small, and the adsorption or migration of individual OH* on the surface or to the interior is constrained by periodic images. Furthermore, at 700K, a clear layer of MgOH/MgO is observed (Figure 9f). Additionally, the diffusion of H* atoms is significantly faster than of OH*, leading to the formation of an oxide species layer on the magnesium surface while H diffuses inward, consistent with observations made using atom probe tomography.[65] At 900K, $H_2O$ in the system is almost completely reacted within the initial 2 ns (Figure 9c). A thicker layer of MgOH/MgO formed on the upper layer of the magnesium substrate (Figure 9g) (in this scenario, the ratio of Mg:$H_2O$ in the model is nearly 1:1). In the figure, OH* represents MgOH, and the difference in consumption between $H_2O$ and MgOH is the quantity of MgO. The ratio of MgOH to MgO here is approximately 8:1.

Subsequently, to prevent the reaction from terminating at the MgOH structure, we increase the proportion of $H_2O$ in the model ($H_2O$:Mg>2) and use a more reactive stepped surface for simulation (Figure 9l). At 700K, the reaction proceeds almost uniformly, and the rate of hydrogen evolution is nearly linear (Figure 9h). Besides, the increase in the thickness of the oxide layer did not significantly inhibit hydrogen evolution. In the final structure of the reaction, pores caused by hydrogen evolution appear in the magnesium substrate, and $H_2$ accumulates in the pores (Figure 9j).

Figure 9m shows the evolution of the pores. As the reaction proceeds, the oxidation/hydroxide layer grows inward and outward, causing local lattice disorder and the formation of cracks. Moreover, the $H_2O$ molecules can penetrate the interior of the oxide layer along the cracks. When the cracks extend into the deeper magnesium substrate, H atoms inside the substrate accumulate at the cracks, and $H_2$ is released along the cracks. Furthermore, the $H_2$ release leads to increased local stress, causing the cracks to expand into pores (as shown in Figure 9j). Finally, when the temperature increases to 900K, the surface is oxidized, reaching equilibrium within 1 ns, and the substrate cracks (Figure 9k). The majority of the magnesium substrate is converted into MgOH.

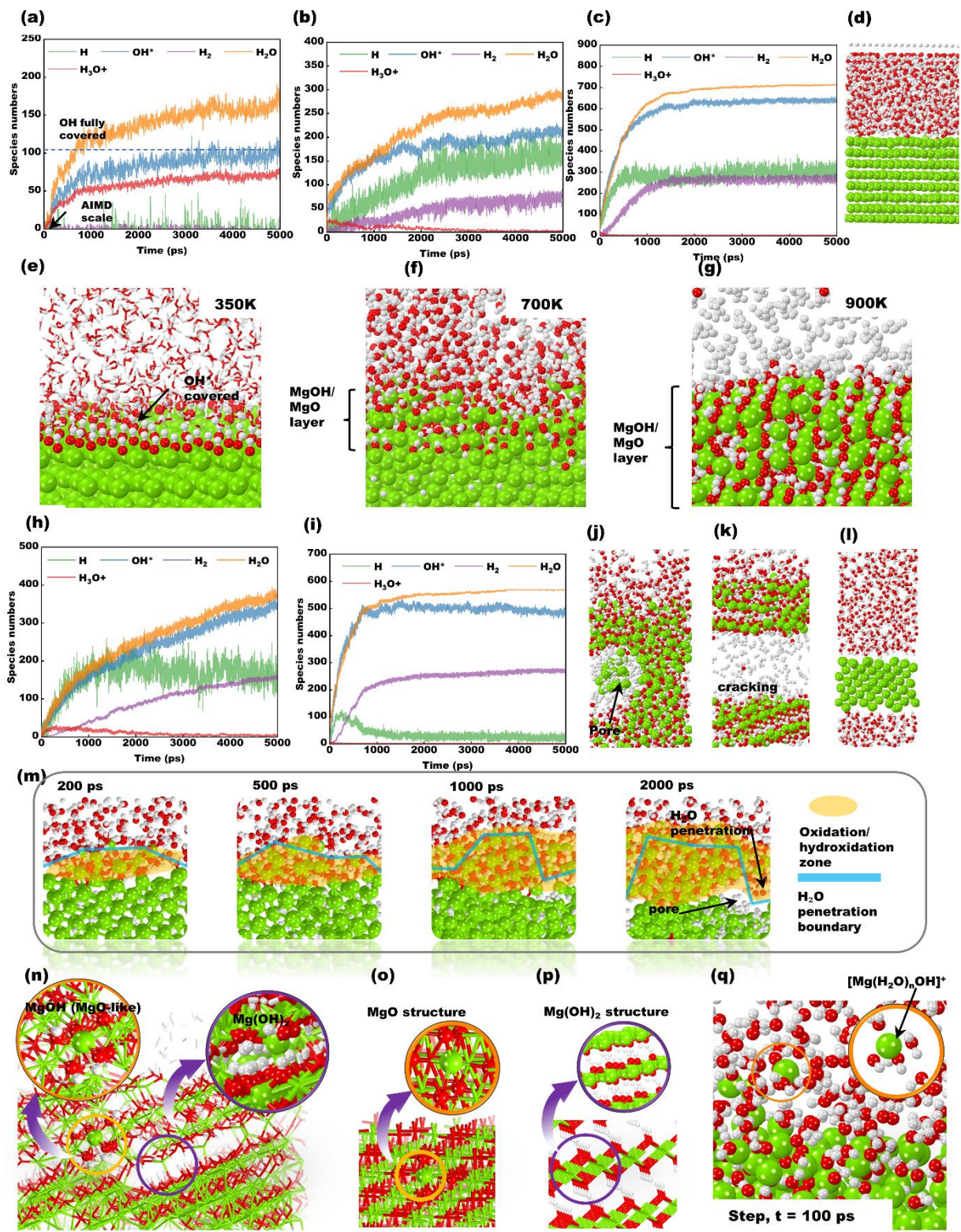

Figure 9. Reaction results for a large system with the close-packed surfaces: species count versus time at (a) 350 K, (b) 700 K, and (c) 900 K, (d) model of the close-packed surface, snapshots of the reaction when (e) t = 5000 ps at 350 K, (f) t = 5000 ps at 700 K, and (g) t = 5000 ps at 900 K; reaction results for the large system with the stepped surfaces: species count versus time at (i) 700 K, (j) 900 K, snapshots of the reaction when (k) t = 5000 ps at 700 K, (l) t = 5000 ps at 900 K, (m) model of the stepped

surface; (n) surface H$_2$O penetrating into the inner layer and forming pores, (o) structure of the reaction product, (p) structure of MgO, (q) structure of Mg(OH)$_2$, and (r) Solvation structure of Mg[(H$_2$O)$_n$OH]$^+$. Atoms are shown as spheres (Mg: green, O: red, H: white); less important atoms are depicted as stick models of the same color.

The structures after the reaction and the related structures used to illustrate the products are shown in Figures 9n-q. Figure 9n shows that MgOH has a MgO-like structure, which exhibits local order and significant crystalline characteristics. In this structure, the central magnesium atom forms ionic bonds with six O atoms, and H is connected to O, with no fixed orientation for H. In fact, due to the low mass of hydrogen, its position is difficult to determine experimentally.

**Comparison with Recent Discoveries.** Recently, researchers have used advanced Cryo-Atom Probe Tomography to analyze the corrosion mechanism of the magnesium surface. The intermediate corrosion phase was determined to be the MgOH phase with a Mg:O:H ratio of 1:1:1.[44] From this, they also deduced two possible structures of the intermediates. The first structure can be found in the inset of Figure 1i. Additionally, the second structure exactly is the reaction product, obtained from our MLMD calculation, which exhibits an MgO-like structure (Figure 9o). In addition, the local Mg(OH)$_2$ structure (Figure 9o) may reveal the production of Mg(OH)$_2$. As MgOH reacts further with H$_2$O, the interlayer spacing between MgOH layers gradually increases, accompanied by the migration of OH, leading to a phase transition towards Mg(OH)$_2$.

Furthermore, recent studies have indicated the possible formation of solvated monovalent magnesium Mg[(H$_2$O)$_n$OH]$^+$ structures on the magnesium surface.[66] In our work, a similar structure is also observed (Figure 9q). The Mg[(H$_2$O)$_n$OH]$^+$ is claimed to reveal the presence of uni-positive Mg$^+$ ions and the absence of protective oxide/hydroxide layers typically formed under anodic/oxidative conditions. However, it must be noted that such direct dissolution of magnesium into solution actually involved 2 electrons lost from the Mg simultaneously in one step, which is believed to be difficult when Song and Atrens proposed that the Mg$^+$ might be involved at film breaks.[67] The 2-electron transfer dissolution may occur under simulated anodic polarization conditions. We acknowledge that this may be a possible mechanism under very strong anodic polarization. Our results without applied potential indicate that magnesium ions can hardly be dissolved directly into the solution. Compared with the direct magnesium dissolution into solution, the formation of protective oxides on magnesium is much more significant. Thus, it can be concluded that without strong anodic polarization, the oxidation of magnesium directly can proceed from the outer layer to the inner layer of the magnesium substrate, forming a protective oxide film, with a small amount of magnesium being dissolved into the electrolyte to form solvated $Mg^{2+}$ or uni-positive $Mg^+$ ions. This is also consistent with the observations by Schwarz et al.[44]

**$Mg(OH)_2$ Layer-Diffusion Barrier and Electrochemical Impact.** To consider the role of the outer $Mg(OH)_2$ layer in electrochemical reactions, we calculate the energy barrier for $Mg^{2+}$ diffusion. Figure 10 reveals that the energy barrier for interlayer diffusion of $Mg^{2+}$ is 2.0 eV, which is greater than the intralayer diffusion value of 1.40 eV. The diffusion barrier for Mg in $Mg(OH)_2$ is relatively high, and as referenced in the Supporting Information (Table S6), this can be considered a difficult diffusion process. Our AIMD calculations for $Mg(OH)_2$ with Mg vacancies have confirmed this viewpoint. To identify potential diffusion pathways, we perform 20 ps AIMD simulations on defective $Mg(OH)_2$ supercells at 1500K, but we do not observe any Mg migration (Supporting Information Figures S13 and S14). Therefore, it can be inferred that in the processes utilizing the electrochemical reactions of magnesium, such as Mg-air batteries, $Mg^{2+}$ primarily diffuses through the electrolyte when migrating from the anode to the cathode, and the presence of $Mg(OH)_2$ hinders this diffusion. Additionally, since $Mg(OH)_2$ is a poor conductor of electrons (Supporting Information Figure S15), even with Mg vacancies, $Mg(OH)_2$ does not exhibit sufficient electronic conductivity, thus it can be considered that the $Mg(OH)_2$ layer impedes the occurrence of electrochemical reactions. Consequently, for Mg-air batteries, the $Mg(OH)_2$ layer can become a hindrance to improving rate performance.[53]

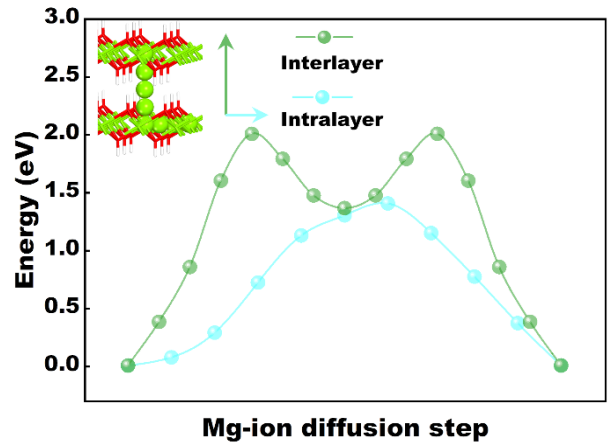

Figure 10. The interlayer and intralayer diffusion energy barrier of Mg ions in Mg(OH)$_2$. The inset in the top-left corner shows the diffusion paths and directions.

**Magnesium Dissolution Mechanism.** After the detailed analyses of the dissolution behavior of magnesium in aqueous media as above, a series of fundamental understandings of the magnesium dissolution process has been revealed. As shown in Figure 11, at the onset of dissolution, the magnesium surface is gradually hydroxylated through processes (a), (b), and (c), with the surface magnesium being covered by OH$_*$. Subsequently, most of the OH$_*$ group directly migrates inward through process (f) to form the MgOH phase, but some of the OH$_*$ group react with hydroxide ions in the solution to form MgO$_*$ (process (e) in Figure 11), and then the adsorbed O$_*$ migrates further inward through process (g) to form MgO. The MgOH phase and MgO together constitute the membrane layer. The MgOH can also react with H$_2$O to produce corrosion products Mg$^{2+}$ and Mg(OH)$_2$, and release anodic hydrogen H$_2$. (not shown in Figure 11). Additionally, as shown in process (h) in the figure, a small portion of OH$_*$ and magnesium ions diffuse into the solution to form solvated Mg[(H$_2$O)$_n$]$^{2+}$ or Mg[(H$_2$O)$_n$OH]$^+$. The H$_3$O$^+$ ions are formed through process (b) and release H$_2$ through a cathodic reaction in process (d). As shown in process (i), the MgOH intermediate phase is solid and directly attached to the magnesium substrate, with a MgO-like structure. The corrosion film grows by extending inward into the magnesium substrate, and this reaction and migration lead to the gradual oxidation of magnesium. After the formation of MgOH, the OH$_*$ further reacts with water to form Mg(OH)$_2$ (as shown in process j). Additionally, as indicated by the water diffusion boundary in Figure 11, the corrosion film is not stable, and H$_2$O can easily penetrate the corrosion membrane layer, leading to the formation of local pores and cracks. The formation of these pores and cracks causes H atoms to accumulate and the evolution of $H_2$ from there.

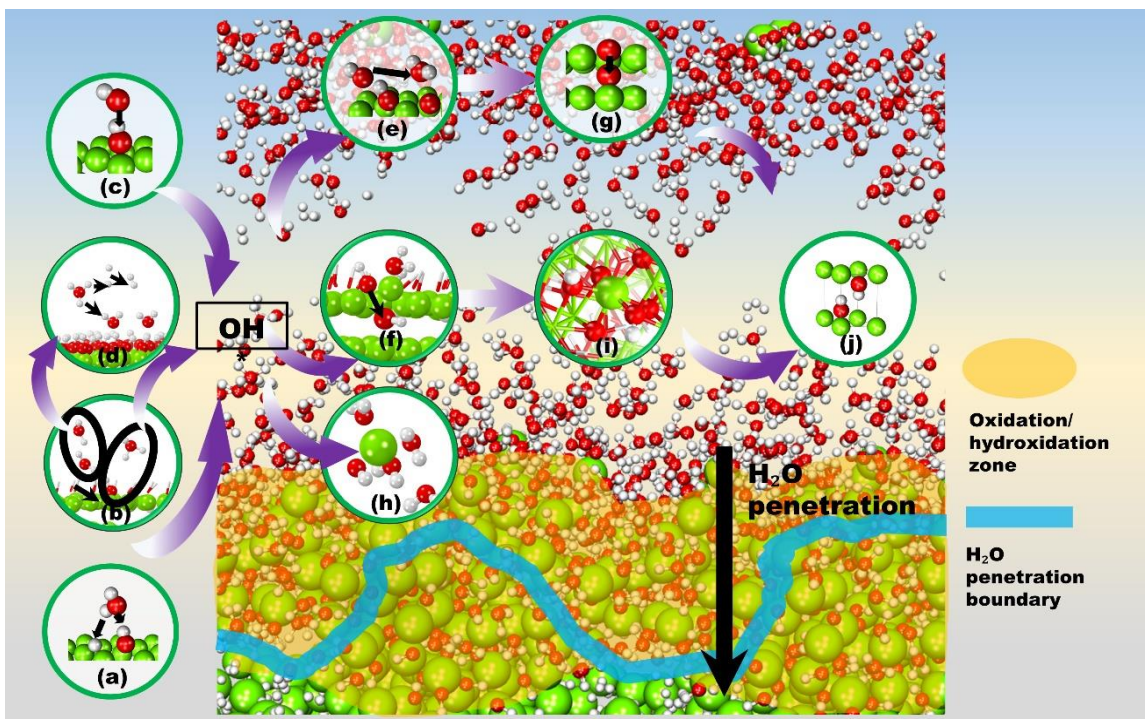

Figure 11. Schematic of the magnesium dissolution mechanism. Atoms are shown as spheres (Mg: green, O: red, H: white).

**Conclusions**

This study elucidates a comprehensive understanding of the dissolution mechanism of magnesium in an aqueous environment based on MLMD simulations and DFT calculations. The NEP-MLP model developed exhibits both high computational accuracy and exceptional efficiency.

AIMD simulations reveal the stepwise hydroxylation of the Mg surface in the initial stages, during which Mg progressively transitions towards a +1 oxidation state. It is interesting that the Mg surface can have a non-integral valence varying from 0 to 2 when exposed in an aqueous solution according to the MLMD simulations and DFT calculations, which very well embodies the involvement of $Mg^+$ in the Mg dissolution process.

During subsequent dissolution, OH∗ and a small fraction of O∗ migrate into the Mg substrate, gradually forming a MgOH/MgO corrosion film primarily composed of MgOH with minor MgO components. This film grows inward into the Mg substrate.

On the Mg surface, in addition to the reaction of water with the MgOH formed in the early stage of Mg oxidation to produce corrosion products and hydrogen, it is also possible that anodic reactions involving $H_2O$ produce MgOH* and $H_3O^+$, while molecular $H_2$ is formed by the recombination of adsorbed hydrogen atoms derived from the reduction of $H_3O^+$. The MgOH is identified as the well-known uni-positive $Mg^+$ species. It exhibits a structure similar to MgO, featuring localized order and distinct crystalline characteristics.

Direct observations of the simulations and calculations reveal the penetration of OH* and $H_2O$ along cracks in the corrosion layer, leading to $H_2$ evolution at these sites. Additionally, solvated species $Mg[(H_2O)_nOH]^+$ plays a minor role in the dissolution process. In an environment with sufficient water, MgOH ultimately transforms into $Mg(OH)_2$, the final corrosion product. The diffusion of $Mg^{2+}$ within the $Mg(OH)_2$ layer is significantly restricted.

This work offers crucial insights into the underlying mechanism of magnesium dissolution and also provides a solid foundation for developing corrosion prevention methods and expanding practical applications of magnesium alloys.

## ACKNOWLEDGMENT

The support of the National Natural Science Foundation of China (No. 52250710159), Shandong Province Excellent Youth Science Fund Project (2023HWYQ-022), Shenzhen Key Laboratory of Advanced Functional Carbon Materials Research and Comprehensive Application (No. ZDSYS20220527171407017) are acknowledged. The scientific calculations in this paper have been done on the HPC Cloud Platform of Shandong University.

Graphical Abstract

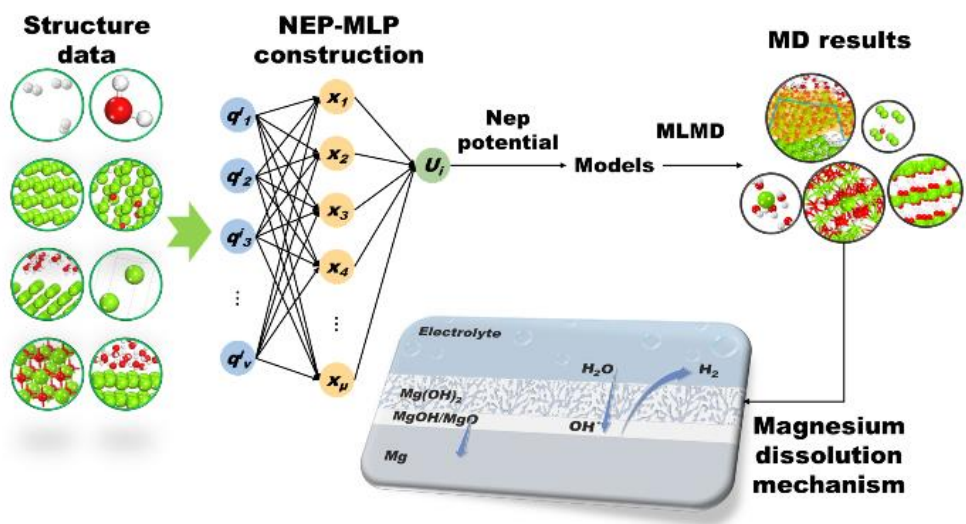